\documentclass[journal]{IEEEtran}
\usepackage[pdftex]{graphicx}

\usepackage{multicol}
\usepackage [autostyle, english = american]{csquotes}

\usepackage{amsmath,amssymb,graphicx}

\MakeOuterQuote{"}
\usepackage[noadjust]{cite}

\begin{document}
\makeatletter
\let\ref\@refstar
\makeatother

\title{Photonic Crystal Microcavities in a\\Microelectronics 45\,nm SOI CMOS Technology}

\author{Christopher V. Poulton, Xiaoge Zeng, Mark T. Wade, Jeffrey M. Shainline, Jason S. Orcutt and Milo\v{s} A. Popovi\'{c}

\thanks{Manuscript received October 28, 2014.}

\thanks{C.V. Poulton, X. Zeng, J.M. Shainline, M.T. Wade, and M.A. Popovi\'{c} are with the Department of Electrical, Computer and Energy Engineering, University of Colorado, Boulder, CO  80309, USA (e-mail: cvpoulton@gmail.com; xiaoge.zeng@colorado.edu; jeffrey.shainline@gmail.com; mark.wade@colorado.edu; milos.popovic@colorado.edu).}
\thanks{J.S. Orcutt is with the Research Laboratory of Electronics, Massachusetts Institute of Technology, Cambridge, MA  02139, USA (email: jsorcutt@mit.edu).}
}

\markboth{IEEE Photonics Technology Letters, vol. X, no. Y, October Z, 2014}{Poulton \MakeLowercase{\textit{et al.}}: Photonic Crystal Microcavities in Zero-change}

\maketitle

\begin{abstract}
We demonstrate the first monolithically integrated linear photonic crystal microcavities in an advanced SOI CMOS microelectronics process (IBM 45\,nm 12SOI) with no in-foundry process modifications.  The cavities were integrated into a standard microelectronics design flow meeting process design rules, and fabricated alongside transistors native to the process. We demonstrate both 1520\,nm wavelength and 1180\,nm cavity designs using different cavity implementations due to design rule constraints.  For the 1520\,nm and 1180\,nm designs, loaded quality factors of 2,000 and 4,000 are measured, and intrinsic quality factors of 100,000 and 60,000 are extracted. We also demonstrate an evanescent coupling geometry which decouples the cavity and waveguide-coupling design.
\end{abstract}


%
\IEEEpeerreviewmaketitle

\section{Introduction}

\IEEEPARstart
{E}{nergy} efficiency and bandwidth density requirements in future CPU-to-memory interconnects have motivated research into monolithic integration of photonics with microelectronics \cite{bajo2009}.  Recent design techniques have enabled photonic devices to be manufactured within standard process design kit (PDK) guidelines in advanced CMOS processes and to be fabricated without requiring any in-foundry process modifications \cite{JasonPhoCMOS,JasonPlat,JasonInte}. Thus, photonics is enabled in native microelectronics fabrication processes, enabling photonics technology to leverage the advances in CMOS technology fabrication at essentially no cost. Nanostructured devices such as photonic crystals (PhCs) require high resolution and low proximity effects which turned previous research in favor of electron beam lithography \cite{LoncarHighQPhC} over photolithography, but this approach is not viable for high volume production. Modern microelectronics CMOS processes, such as the 45\,nm process used in this work, support the resolution and process control to define PhCs and provide a scalable solution for mass manufacturing. However, they are entirely optimized for electronic circuits with no provision for photonics.  PhC microcavities are potential building blocks for efficient filtering, tuning, modulation, all optical switching and nonlinear applications \cite{Notomi,LoncarPhCThermal}. Therefore, their direct integration into advanced CMOS, alongside state-of-the-art microelectronics, may impact the commercial viability of electronic-photonic systems in a number of applications.

We demonstrate efficient linear photonic crystal cavities in a state-of-the-art microelectronics CMOS process -- a silicon-on-insulator (SOI) CMOS transistor process -- implemented in the transistor device body layer.  We demonstrate 1520\,nm design devices with a loaded quality factor of 2,150 (92 GHz bandwidth), and extract an intrinsic quality factor on the order of 100,000. Cavities with a 1180\,nm resonant design wavelength with an extracted intrinsic quality factor on the order 60,000 are also presented. All cavities are excited via evanescent coupling \cite{manolatou}, enabling decoupled design of the microcavity and waveguide coupling [Fig.~\ref{fig:stackup}(a)].

\section{Advanced CMOS Design Considerations}

We employ the IBM 45\,nm 12SOI process \cite{ibm12soi} to fabricate the devices.  Recent work has demonstrated linear PhC cavities in a bulk silicon (polycrystalline transistor-gate device layer) process \cite{BulkSiPhC}, also promising for CMOS integration.  An advantage of an SOI CMOS process, in comparison to bulk CMOS, is the low optical loss of the crystalline silicon transistor body layer when used as the waveguiding layer.  The cross-section of the cavity within the 12SOI process is illustrated in Fig.~\ref{fig:stackup}(b) (exact layer thicknesses available in IBM 12SOI Process Design Kit under NDA \cite{ibm12soi}), showing the body silicon  layer waveguide, and a nitride stressor layer above it (present in advanced-node processes to increase mobility in the MOSFET channel region).  The buried oxide is not thick enough to enable optical isolation between the waveguides and the silicon substrate. Thus, a post-processing XeF$_2$ silicon etch step is required to remove the silicon substrate. The substrate removal can be performed locally \cite{HolzwarthCLEO2008} or globally \cite{JasonPhoCMOS}, and it was previously shown that the substrate removal does not degrade transistor performance \cite{JasonPlat}. The experiments in this work utilized global substrate removal. A micrograph image of a PhC after substrate removal is shown in Fig.~\ref{fig:simulations}(c).

The primary challenges in design are the sub-90\,nm thickness of the body silicon layer which limits confinement, and process design rules including minimum feature size, enclosed area and notch width rules.  Because of the thin body silicon layer, the cavity waveguide width is large relative to that of typical silicon designs \cite{LoncarHighQPhC} in order to maximize confinement [Fig.~\ref{fig:stackup}(c)], and the cavity is also longer to support a high intrinsic quality factor.  Although the polysilicon gate layer could be utilized on top of the crystalline silicon body layer to increase confinement \cite{JasonInte}, it is omitted here because its substantial optical loss would degrade the intrinsic quality factor.  In order to pass to the fabrication stage within a standard microelectronics process, these structures must pass automatic design rule checks (DRC) provided for the process.  Unit cells with discretized circular holes are prone to violation of notch design rules.  As a result, the 1520\,nm resonant cavities presented here use square holes in silicon as the unit cells to simplify layout and DRC conformance. The 12SOI process has a relatively large minimum enclosed area rule which places a strong constraint on the cavity design. Therefore, in cavities designed for 1180\,nm wavelength, an alternative unit cell design of isolated rectangles of body silicon was used. Figures~\ref{fig:stackup}(a) and (e) show the layout for the 1520\,nm and 1180\,nm designs, respectively.

\section{Cavity and Coupling Design}
\label{sec:Design}

\begin{figure}[t]
\centerline{\includegraphics[width=.8\columnwidth]{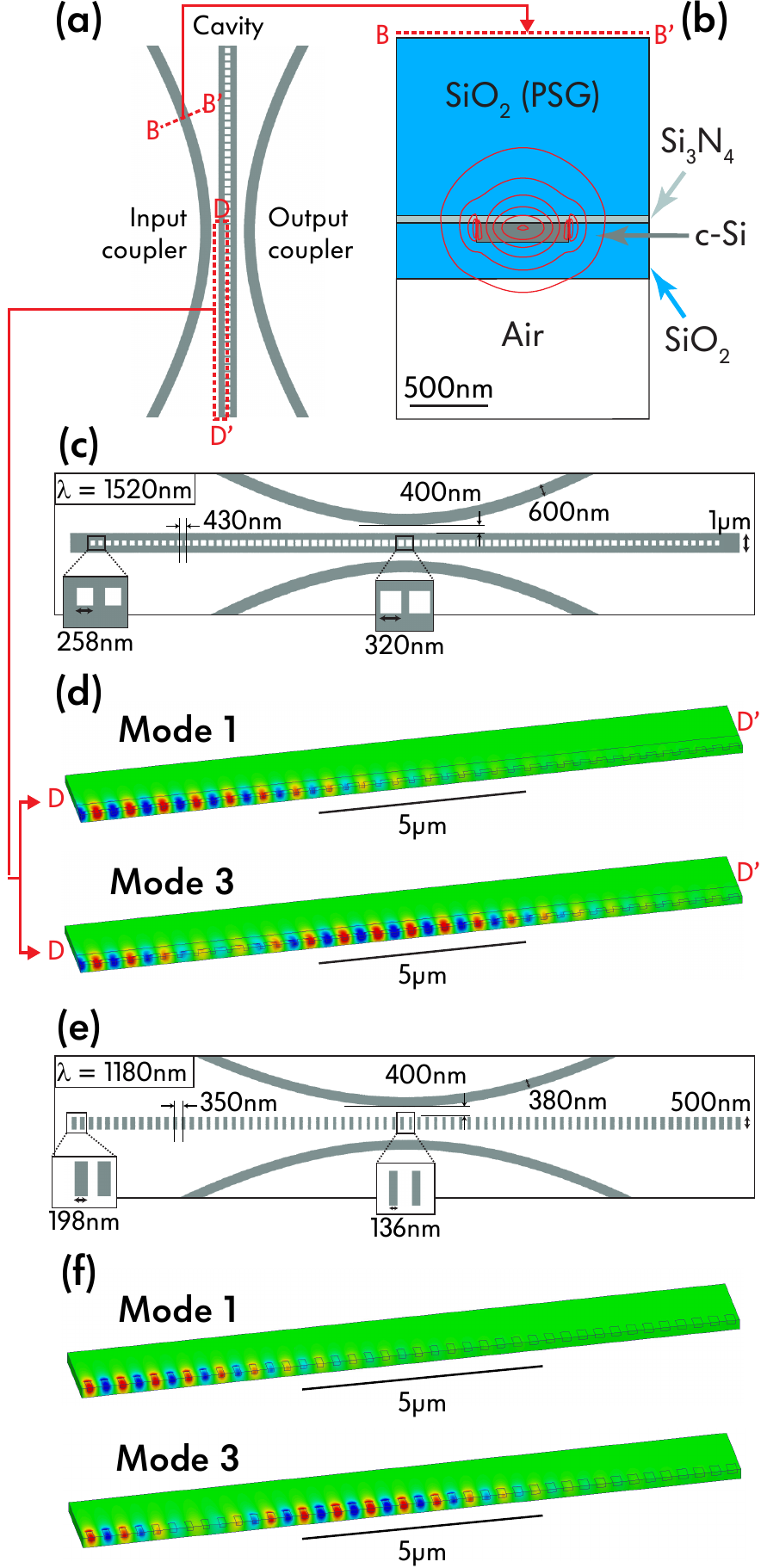}} 
\caption{(a) Device geometry showing the photonic crystal cavity and evanescently coupled input and output waveguides; (b) waveguide cross-section in the IBM 45\,nm 12SOI CMOS process showing the fundamental TE mode; (c) 1520\,nm device design with (d) first and third longitudinal modes; (e) 1180\,nm device design (unit cell difference due to design rules) with (e) first and (f) third longitudinal modes. \label{fig:stackup}}
\vspace{-8pt}
\end{figure}

\begin{figure}[t!]
\centerline{\includegraphics[width=.8\columnwidth]{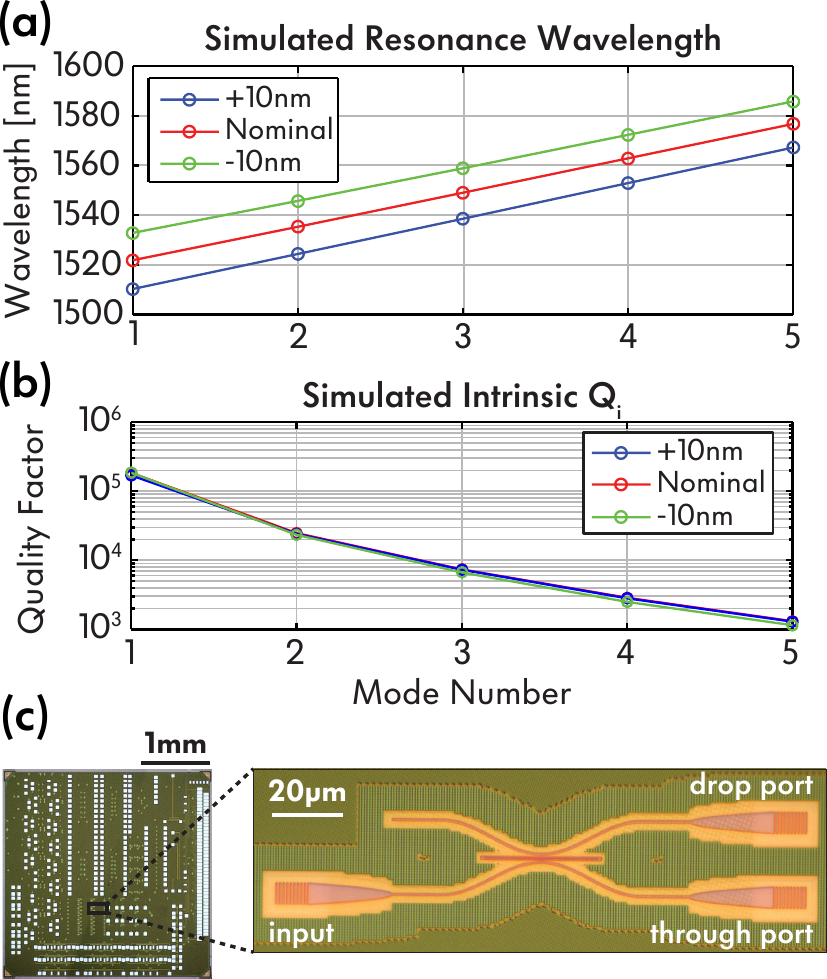}}
\caption{(a) Simulated resonance wavelengths of the 1520\,nm cavity for the nominal design and $\pm10$\,nm hole size design variants; (b) simulated intrinsic loss $Q_i$'s; (c) optical micrographs of the 3$\times$3\,mm chip and of a fabricated device including grating coupler ports, waveguides and a cavity. \label{fig:simulations}}
\vspace{-8pt}
\end{figure}

The cavities are synthesized to support Hermite-Gaussian resonant modes, i.e. to approximate a truncated parabolic optical potential \cite{LoncarPhCDesign}.  In contrast to analytic approximations typically used, we employed a synthesis procedure that relies on a parameter map obtained from rigorous 3D numerical band-structure calculations.  High-Frequency Structure Solver (HFSS) \cite{AnsoftHFSS} was used for photonics simulations.  First, it was used in eigensolver configuration with periodic longitudinal boundary conditions to compute the photonic band structure.  This provided the mirror strength of the cavity unit cells at a target resonance wavelength as a function of a unit cell geometry parameter, such as the size of the square holes or rectangular blocks. A cavity design (with a non-uniform cell distribution) was synthesized from the parameter maps. HFSS was used to find the fundamental and higher order modes of the full cavity to verify the synthesis procedure, and to analyze effects of fabrication variations. The synthesized 1520\,nm design cavity has a single transverse mode and multiple longitudinal modes [Fig.~\ref{fig:stackup}(d)], and the simulated free spectral range (FSR) is 1.71\,THz.  Fig.~\ref{fig:simulations}(a) shows the simulated resonant wavelength for each mode and Fig.~\ref{fig:simulations}(b) shows the simulated intrinsic quality factors, $Q_i$.  The fundamental mode of the nominal design has a simulated $Q_i$ of 184k at 1521\,nm. $Q_i$ decreases exponentially with mode number because the cavity length is fixed, the higher order longitudinal modes occupy a larger portion of the cavity which results in scattering due to the finite extent of the cavity.

The cavity is excited via evanescent coupling from two side-coupled waveguides, an input and a drop waveguide, in a symmetric configuration [Fig.~\ref{fig:simulations}(c)].  In comparison to direct excitation from the waveguide in which the cavity is formed, such a coupling geometry has the advantage that the cavity design is independent of coupling design; the cavity-waveguide gap is the only parameter changed when adjusting coupling, and no cavity redesign is needed.  Assuming equal coupling from both side-coupled waveguides, a single external quality factor can be defined that relates the intrinsic loss quality factor and the total (loaded) quality factor. The coupled-mode theory model relating these quality factors and the through port transmission on resonance is \cite{manolatou}:

\vspace{-10pt}
\begin{align}
\frac{1}{Q_t} &= \frac{1}{Q_i}+\frac{1}{Q_e}\label{equ:invQs}\\
\frac{P_{thru}}{P_{in}} &= \left(\frac{1+2 Q_e/Q_i}{2+2 Q_e/Q_i}\right)^2\label{equ:CMTmodel}
\end{align}

\noindent where $Q_e$, $Q_i$ and $Q_t$ are the external, intrinsic and total (loaded) quality factors, and, $P_{thru}$ and $P_{in}$ are the optical powers in the through port and input port. In this configuration, the ideal transmission on resonance is $-6$\,dB (25\%) to all four ports due to the symmetry of the standing-wave cavity system.

\section{Measurements and Analysis}

\begin{figure}[t]
\centerline{\includegraphics[width=.8\columnwidth]{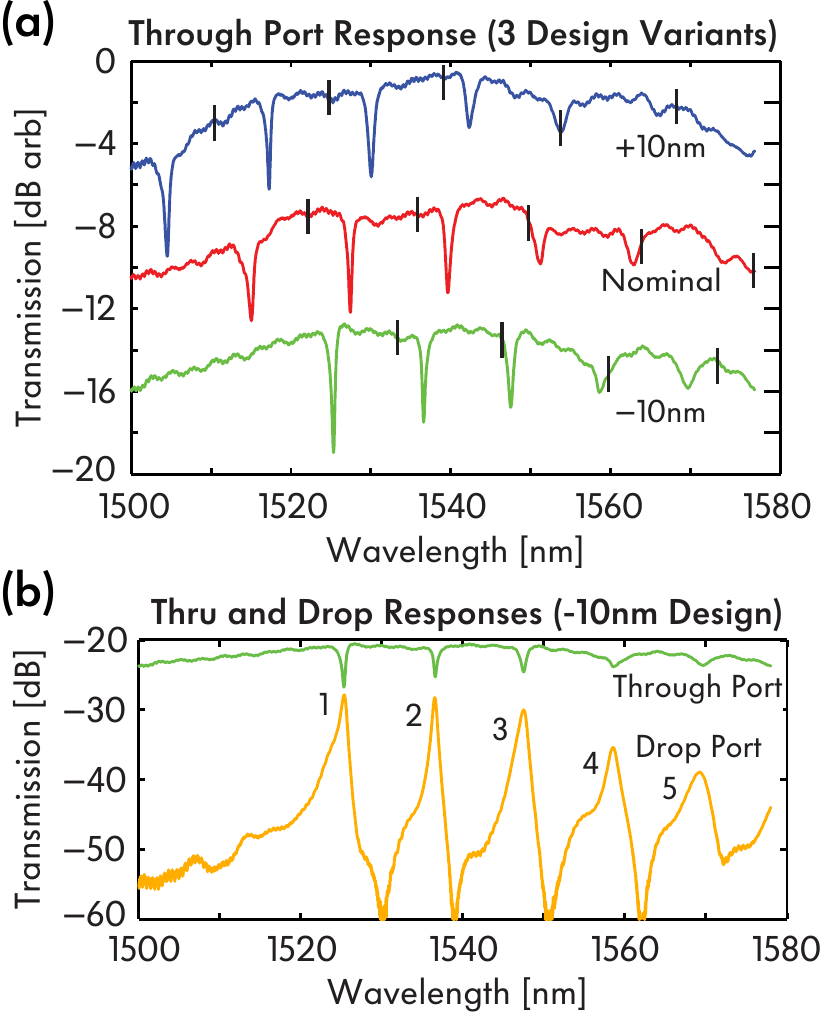}}
\caption{(a) Cavity through port responses (nominal and $\pm$10\,nm hole size), vertical lines show design resonances; (b) through and drop port responses with longitudinal mode numbers labeled ($-10$\,nm hole design).\label{fig:spectra}}
\vspace{-8pt}
\end{figure}

\begin{figure}[t!]
\centerline{\includegraphics[width=.8\columnwidth]{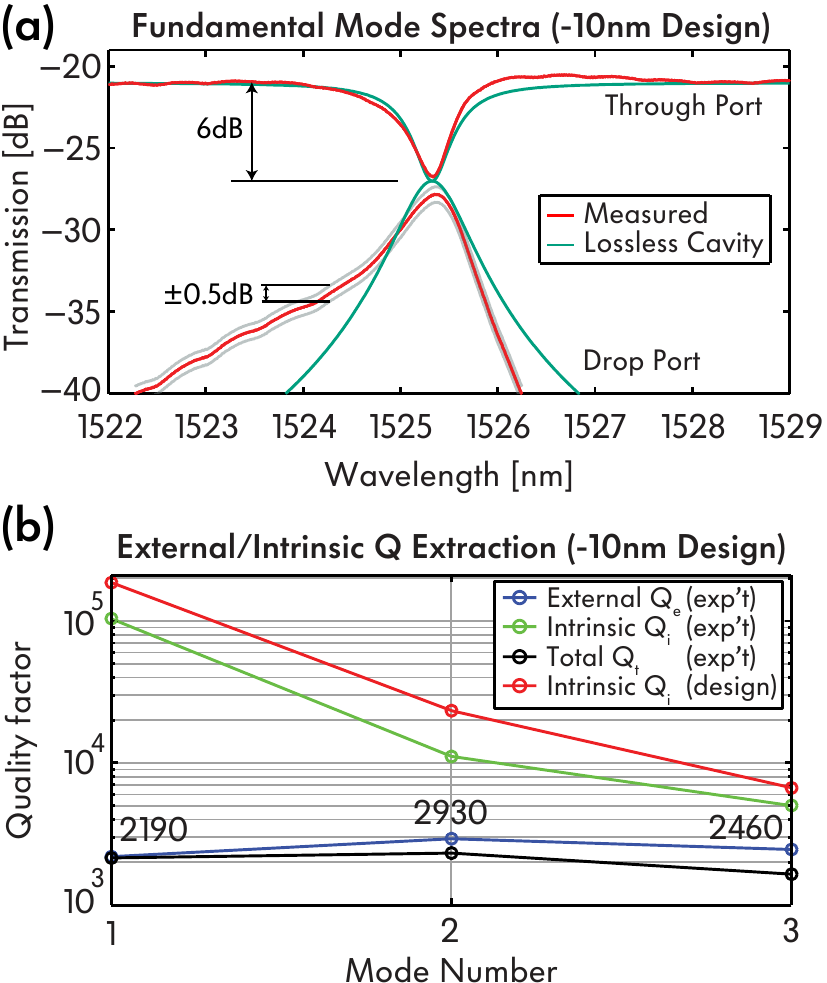}}
\caption{(a) Fundamental mode of $-10$\,nm design cavity along with an ideal response; (b) measured total, external and intrinsic $Q$'s, and design intrinsic $Q_i$ of the first three modes for the  $-10$\,nm design cavity.\label{fig:extractedq}}
\vspace{-8pt}
\end{figure}

Fig.~\ref{fig:spectra}(a) shows measured through port responses of three 1520\,nm design cavities -- the nominal design and variants with $\pm$10\,nm square hole side dimensions to account for process variability -- and the simulated design resonances. Measured resonances are about 7\,nm away from design which is expected due to fabrication variation in the device layer thickness.  The fundamental through the $5^{\textrm{th}}$ order longitudinal modes are seen in the transmission spectra. The nominal design shows an FSR of about 1.52\,THz which is close to the design FSR of 1.71\,THz. Fig.~\ref{fig:spectra}(b) shows the through and drop port spectra of the $-10$\,nm cavity design and Fig.~\ref{fig:extractedq}(a) shows a close-up view of the fundamental mode which shows transmission near the ideal value of $-6$\,dB, i.e. the device operates as a wavelength selective 4-way power splitter.  The bandwidth is 92\,GHz (a $Q_t$ of 2,150). 

Eq.~(\ref{equ:invQs}) and Eq.~(\ref{equ:CMTmodel}) can be used to extract the external quality factor, $Q_e$, as well as the intrinsic quality factor, $Q_i$, of the cavity for each mode. The on-resonance transmission and bandwidth in the through port response alone provide all the coupling parameters. Transmission near $-6$\,dB  indicates that most power is coupled to ports and that the total quality factor, $Q_t$, is dominated by $Q_e$ due to external coupling to the waveguides. Therefore, parameter extraction to find the $Q_i$ of the cavity is sensitive to errors -- for this $Q_t$, an extinction larger than 5.3\,dB  ensures $Q_i>$ 25,000, and an extinction larger than 5.84\,dB ensures a $Q_i>$ 100,000. Measured spectra give extinction of 5.94\,dB when normalizing out the grating coupler response.  We estimate an uncertainty of about 0.1\,dB due to Fabry-Perot oscillations, so $Q_i$ is on the order of 100,000. Fig.~\ref{fig:extractedq}(b) shows the extracted $Q_e$'s and $Q_i$'s of the $-10$\,nm design cavity for the first three modes.

Another observation in this coupling geometry is that the extracted $Q_e$ is higher for the second mode than for both the first and third mode [Fig.~\ref{fig:extractedq}(b)].  This is consistent with expected behavior. Even numbered modes have a field null in the center of the cavity where the input and output waveguides are located which suppresses the coupling. This means that even modes have a higher $Q_e$ compared to the following odd longitudinal mode. This is directly measured by a greater $Q_t$ for the second mode in spectra in all cavity designs and confirmed by HFSS simulations with a coupling bus included.


\begin{figure}[t!]
\centerline{\includegraphics[width=.8\columnwidth]{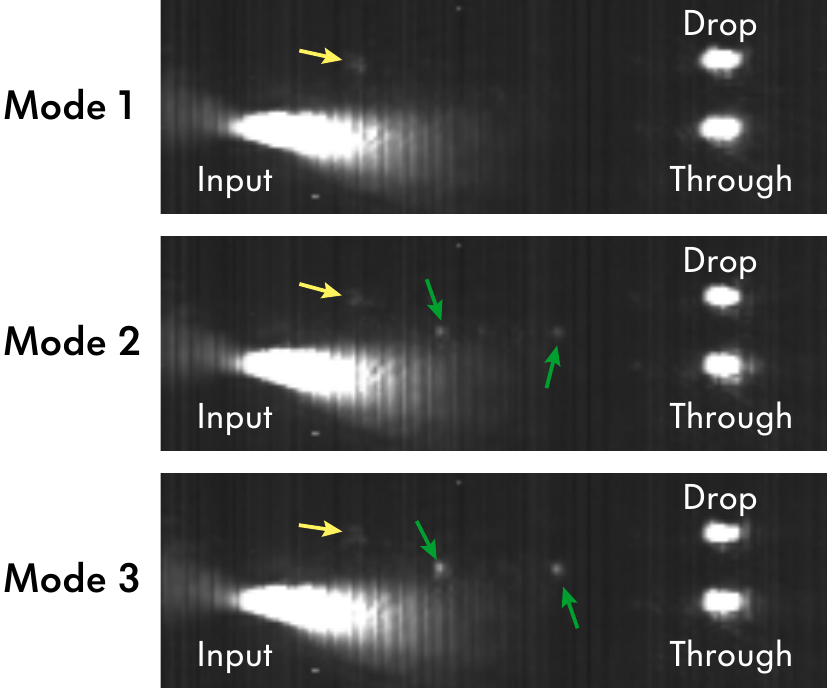}}
\caption{Top view IR images of 1520\,nm design cavity on resonance. Arrows point to scattering at the shunt waveguide (yellow) and on the edges of the cavity nanobeam (green).\label{fig:ircam}}
\end{figure}

\begin{figure}[t!]
\centerline{\includegraphics[width=.8\columnwidth]{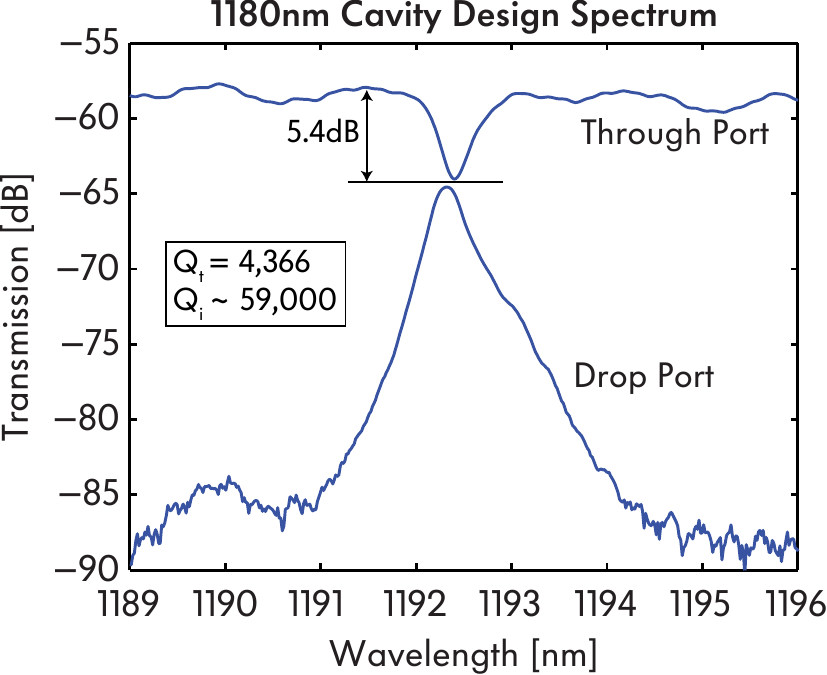}}
\caption{Through and drop port response of the 1180\,nm fundamental wavelength design cavity. A total $Q_t$ of 4,366 is measured with an estimated intrinsic $Q_i$ of 59,000.\label{fig:1180nmCavs}}
\vspace{-8pt}
\end{figure}

We note that $Q_i$ exponentially drops with mode number and explain this as a tunneling to the guided nanobeam waveguide mode at the edges of the cavity.  This is confirmed by IR images [Fig.~\ref{fig:ircam}] when each of the first three modes was excited on resonance.  In all three cases, scattering is seen at the terminated waveguide port as well as the through and drop port grating couplers, as expected for a standing wave cavity that radiates to all four ports.  The cavity is dark on resonance which is a good indicator that radiation loss, at least within the NA of the microscope objective, is small and is consistent with high $Q_i$. On the two higher order mode images, some scattering is seen also on the edges of the cavity itself.  This is consistent with increased tunneling radiation loss for the higher order modes.

Cavities were also synthesized using the same process given in Section~\ref{sec:Design} with a 1180\,nm design resonance wavelength [Fig.~\ref{fig:stackup}(e, f)]. Designs at 1180\,nm wavelength are of interest due to compatibility with Si-Ge photodetectors, created within the same SOI CMOS process \cite{1180nmPDs}, which have higher absorption at lower wavelengths.  To enable designs at 1180\,nm that conform to the design rule constraints, the unit cell of the cavities was an isolated silicon rectangular block.  The fundamental mode of this design has a simulated intrinsic quality factor of 845k near 1180\,nm.  Fig.~\ref{fig:1180nmCavs} shows the measured fundamental mode transmission.  The bandwidth is 57.6\,GHz (a $Q_t$ of 4,366).  After fitting and removing the large Fabry-Perot oscillations, the device has an extinction ratio of about 5.4\,dB and an extracted $Q_i$ of about 59,000.

\section{Conclusion}
We demonstrated the first linear photonic crystal microcavities in an advanced CMOS microelectronics process. With intrinsic quality factors of $\sim$100k and $\sim$60k for 1520\,nm and 1180\,nm designs, respectively, these cavities offer potential solutions for both passive (e.g. filtering and power splitting) and active (e.g. modulation and detection) applications. Active designs are enabled by utilizing p-type and n-type implants that already exist in the process and are traditionally used for transistors. Silicon germanium (used for transistor strain engineering) can be used to enable detector designs.

\section*{Acknowledgment}

This work was supported by DARPA POEM program award HR0011-11-C-0100.


\begin{thebibliography}{99}


\bibitem{bajo2009} C. Batten, A. Joshi, J. Orcutt, A. Khilo, B. Moss, C.W. Holzwarth, M.A. Popovi\'{c}, H. Li, H.I. Smith, J.L. Hoyt, F.X. K\"{a}rtner, R.J. Ram, V. Stojanovi\'{c} and K. Asanovi\'{c}, ``Building Many-Core Processor-to-DRAM Networks with Monolithic CMOS Silicon Photonics,'' \textit{IEEE Micro}, vol. 29, no. 4, pp. 8-12, Jul./Aug. 2009.

\bibitem{JasonPhoCMOS} J.S. Orcutt and R.J. Ram, ``Photonic Device Layout Within the Foundry CMOS Design Environment,'' \textit{IEEE Photon. Technol. Lett.}, vol. 22, no. 8, pp. 544-546, Apr. 2010.

\bibitem{JasonPlat} J.S. Orcutt, B. Moss, C. Sun, J. Leu, M. Georgas, J. Shainline, E. Zgraggen, H. Li, J. Sun, M. Weaver, S. Uro\u{s}evi\'{c}, M. Popovi\'{c}, R.J. Ram and V. Stojanovi\'{c}, ``Open foundry platform for high-performance electronic-photonic integration,'' \textit{Opt. Express}, vol. 20, no. 11, pp. 12222-12232, May 2012.

\bibitem{JasonInte} J.S. Orcutt, A. Khilo, C.W. Holzwarth, M.A. Popovi\'{c}, H. Li, J. Sun, T. Bonifield, R. Hollingsworth, F.X. K\"{a}rtner, H.I. Smith, V. Stojanovi\'{c} and R.J. Ram, ``Nanophotonic integration in state-of-the-art CMOS foundries,'' \textit{Opt. Express}, vol. 19, no. 3, pp. 2335-2346, Jan. 2011. 

\bibitem{LoncarHighQPhC} P.B. Deotare, M.W. McCutcheon, I.W. Frank, M. Khan and M. Lon\v{c}ar, ``High quality factor photonic crystal nanobeam cavities,'' \textit{Appl. Phys. Lett.}, vol. 94, no. 12, p. 121106, Mar. 2009.


\bibitem{LoncarPhCThermal} P.B. Deotare, L.C. Kogos, I. Bulu and M. Lon\v{c}ar, ``Photonic crystal nanobeam cavities for tunable filter and router applications,'' \textit{IEEE Journal of Selected Topics in Quantum Electronics}, vol. 19, no. 2, p. 3600210 Mar./Apr. 2013.

\bibitem{Notomi} M. Notomi, A. Shinya, K. Nozaki, T. Tanabe, S. Matsuo, E. Kuramochi, T. Sato, H. Taniyama and H. Sumikura, ``Low-power nanophotonic devices based on photonic crystals towards dense photonic network on chip,'' \textit{IET Circuits, Devices \& Systems}, vol. 5, no. 2, pp. 84-93, Mar. 2011.

\bibitem{manolatou} C. Manolatou, M.J. Khan, S. Fan, P.R. Villeneuve, H.A. Haus and J.D. Joannopoulos, ``Coupling of Modes Analysis of Resonant Channel Add-Drop Filters,'' \textit{IEEE Journal of Quantum Electronics}, vol. 35, no. 9, pp. 1322-1331, Sep. 1999.

\bibitem{ibm12soi} S. Lee, B. Jagannathan, S. Narasimha, A. Chou, N. Zamdmer, J. Johnson, R. Williams, L. Wagner, J. Kim, J.-O. Plouchart, J. Pekarik, S. Springer, and G. Freeman, ``Record RF performance of 45-nm SOI CMOS Technology,'' in \textit{IEDM Dig. Tech. Papers}, 2007, pp. 255-258.

\bibitem{BulkSiPhC} K.K. Mehta, J.S. Orcutt and R.J. Ram, ``Fano line shapes in transmission spectra of silicon photonic crystal resonators,'' \textit{Appl. Phys. Lett.}, vol. 102, no. 8, p. 081109, Feb. 2013.

\bibitem{HolzwarthCLEO2008} C.W. Holzwarth, J.S. Orcutt, H. Li, M.A. Popovi\'{c}, V.M. Stojanovi\'{c}, J.L. Hoyt, R.J. Ram and H.I. Smith, ``Localized Substrate Removal Technique Enabling Strong-Confinement Microphotonics in Bulk Si CMOS Processes,'' in \textit{Conference on Lasers and Electro-Optics}, OSA Technical Digest (CD) (Optical Society of America, 2008), paper CThKK5. 

\bibitem{LoncarPhCDesign} Q. Quan and M. Lon\v{c}ar, ``Deterministic design of wavelength scale, ultra-high Q photonic crystal nanobeam cavities,'' \textit{Opt. Express}, vol. 19, no. 19, pp. 18529-18542, Sep. 2011.

\bibitem{AnsoftHFSS} ANSYS HFSS, Release 14.0.0.

\bibitem{1180nmPDs} M. Georgas, B.R. Moss, C. Sun, J. Shainline, J.S. Orcutt, M. Wade, Y.-H. Chen, K. Nammari, J.C. Leu, A. Srinivasan, R.J. Ram, M.A. Popovi\'{c} and V. Stojanovi\'{c}, ``A Monolithically-Integrated Optical Transmitter and Receiver in a Zero-Change 45nm SOI Process,'' in \textit{Proc. VLSI Symposia}, 2014, paper C6p4.

\end{thebibliography}
\end{document}